\documentclass[%
 reprint,
 amsmath,amssymb,
 aps,
]{revtex4-2}

\usepackage{graphicx}% Include figure files
\usepackage{dcolumn}% Align table columns on decimal point
\usepackage{bm}% bold math
\usepackage{lineno}
\usepackage{siunitx}
\begin{document}

\preprint{APS/123-QED}

\title{Imaging of high-frequency electromagnetic field by multipulse sensing using nitrogen vacancy centers in diamond}% Force line breaks with \\

\affiliation{
Division of Physics, Univ. of Tsukuba, 1-1-1 Tennodai, Tsukuba, Ibaraki, 305-8571, Japan
}
\author{Shintaro Nomura}
 \email{nomura.shintaro.ge@u.tsukuba.ac.jp}
\affiliation{
Division of Physics, Univ. of Tsukuba, 1-1-1 Tennodai, Tsukuba, Ibaraki, 305-8571, Japan
}

\author{Hideyuki Watanabe}%
\affiliation{
National Institute of Advanced Industrial Science and Technology (AIST) Central2, Umezono, Tsukuba, Ibaraki, 305-8568, Japan
}

\author{Satoshi Kashiwaya}%
\affiliation{
Department of Applied Physics, Nagoya Univ. Chikusa-Ku, Nagoya, Aichi, 464-8571, Japan
}

%\affiliation{%
\begin{abstract} 
Near-field enhancement of the microwave field is applied for imaging high frequency radio field using a diamond chip with an $n$-doped isotopically purified diamond layer grown by microwave plasma assisted chemical vapor deposition. A short $\pi$ pulse length enables us to utilize a multipulse dynamic decoupling method for detection of radio frequency field at 19.23 MHz. An extraordinary  frequency resolution of the external magnetic field detection is achieved by using amplitude-shaped control pulses. Our method opens up the possibility for high-frequency-resolution RF imaging at $\mu$m spatial resolution using nitrogen vacancy centers in diamond.
\end{abstract}

\maketitle
\section{Introduction}

Electromagnetic field imaging has been widely utilized in material characterizations, characterization of radio frequency (RF) devices, and clinical applications~\cite{Zoughi00}. Antennas and coils are commonly used to detect RF band electromagnetic waves. A spatial resolution has been limited by sensor size, except for imaging methods using strong gradient magnetic fields and Fourier transform in magnetic resonance imaging. A diamond nitrogen vacancy (NV) center, which consists of a nitrogen impurity and an adjacent vacancy with a N-V distance of 0.169 nm in diamond, is used as a very small sensor. Since RF sensors using NV centers enable potentially high spatial resolution imaging, they have been actively studied recently.~\cite{Chipaux15, Horsley18, Yang18, Giacomo20, Hu19, Mizuno20, Nomura21} By using diamond NV centers, the external electromagnetic field distribution is transferred to NV electron spin states using quantum sensing techniques. Then NV electron spin states are read from photoluminescence from NV centers. A high-throughput method using optical microscopy is widely used to obtain RF field distribution, whose spatial resolution is determined by the optical diffraction limit.

Methods to achieve high sensitivity have been studied so far by decoupling from environmental noise. 
Two methods have been applied to high-sensitivity imaging of radio-frequency electromagnetic fields. The first method is to detect RF field through spin locking~\cite{Loretz13, Ohashi13, Rosskopf14, Nomura21} In this method, the level-splitting by the dressed state under intense resonant microwave driving is utilized to detect RF fields in high sensitivity. The splitting energy is set by the Rabi frequency of the microwave driving field ($\Omega$). The range of the RF frequency to be detected is set by the Rabi frequency of the microwave field. High Rabi frequency up to about 1 GHz is available in diamond NV centers ~\cite{Fuchs09}. Thus the spin-locking method may potentially be applied to detect electromagnetic fields in the wide range of 1 MHz – 1 GHz. A drawback of this method is that the spectral resolution of the RF field detection is ultimately limited by the ${T_1}$ time of NV centers, which is several ms at room temperature. 

The second method is to utilize dynamic decoupling to reduce disturbance from noise in the environment. 
By applying many  $\pi$ pulses at an interval of $\tau$, the signal at the frequency $f_{0}$ = $\frac{1}{2\tau}$ is selectively detected, while noise away from $f_0$ is filtered out. Frequency selectivity is controlled by the number of  $\pi$ pulses. CPMG and XY8 methods are commonly used for dynamic decoupling.~\cite{Cywinski08, Ryan10, Lange10, Yuge11, Wang12}
The pulse sequence of the XY8 method is designed to compensate pulse errors, which is a desirable feature for imaging.
The range of the RF frequency to be detected is set by the pulse interval of $\tau$, and hence, is limited by the $\pi$  pulse length. For the detection of higher frequency electromagnetic fields, the spin-locking method is more suitable. Application of the dynamic decoupling method to high RF frequency remains  few.~\cite{Zopes17}

High frequency ac field detection at the frequency close to the driven transition $\omega_0$ has been demonstrated.~\cite{Joas2017, Stark2017, Saijo2018, Tashima19} This method up-converts the detection frequency to $\omega$ = $\omega_0 \pm \Omega$ for spin locking and $\omega$ = $\omega_0 \pm \pi$/$\tau$ for dynamic decoupling,~\cite{Joas2017} and enables detection of GHz microwave at high sensitivity. A method is also developed to use two driving fields of the Rabi frequencies $\Omega_1$ and $\Omega_2$ to detect signal frequency $\omega$ = $\omega_0$ + $\Omega_1$ + $\Omega_2$/2.~\cite{Stark2017}
Nevertheless, the range of the detection frequency is limited by the amplitude of the driving field. For example, there remains a frequency region, $\Omega < \omega < \omega_0 - \Omega$ or $\pi$/$\tau < \omega < \omega_0 - \pi$/$\tau $, inaccessible by the method by Joas et al.~\cite{Joas2017} Filling this gap is important in some applications where the driven frequency $\omega_0$ cannot be tuned independent of the detection frequency $\omega$. One such important application is NMR spectroscopy, where the $^{1}$H NMR frequency is set by the external bias magnetic field and hence, by $\omega_0$. For example, the bias magnetic field is set to 0.5 T to detect 21 MHz $^{1}$H NMR resonance at $\omega_0$/$(2\pi)$ = 11 GHz, where the above methods fail to be applied unless sufficiently large $\Omega$ or short $\tau$ is prepared. Consequently, an extension of $\Omega$ and $\pi$/$\tau$ is highly desired. 

We previously showed that the Rabi frequency over a thin metal wire structure was locally enhanced compared to the Rabi frequency away from the metal wire due to the near-field enhancement of the microwave field in the vicinity of a thin wire.~\cite{Giacomo20} This method enables local driving of the electron spins of NV centers and thus the application of microwave pulses with short $\pi$ pulse length with minimum unwanted heat generation by the microwave field.  In this pper, we utilize the near-field enhancement of the miacrowave field to obtain a short $\pi$  pulse length using an intense microwave field of the Rabi frequency at 40 MHz, and demonstrate RF field imaging at 19.23 MHz using the XY8 method, thus extending the detection range of the dynamical decoupling method to higher frequency.

\section{Experimental}

A 300 nm-thick $^{15}$N-doped, isotopically purified diamond $^{12}$C layer was grown on a (100)-oriented ultra-pure diamond chip (Element Six Ltd., electronic grade) by microwave plasma assisted chemical vapor deposition (CVD)~\cite{Ishikawa12, Ohashi13, Zeeshawn21} from isotopically enriched $^{12}$CH$_{4}$ (99.999\% for $^{12}$C), H$_{2}$ and isotopically enriched $^{15}$N$_{2}$ mixed gas. The diamond chip has dimensions of $2.0 \times 2.0$ mm$^{2}$ and a thickness of 0.5 mm. Nitrogen gas was introduced to intentionally create NV centers. 
After the deposition of a $^{15}$N-doped diamond layer, 
He$^{+}$ ions were implanted to enhance NV$^{-}$ conversion. Following the ion implantation step, the diamond chip was annealed in Ar/H$_2$ gas at 1000$^{\circ}$C for 1.5 h, followed by annealing in air at 450$^{\circ}$C for 24 h.~\cite{Yamamoto13,Zeeshawn21}

A Ti/Au wire with a width of  10 $\mu$m as schematically shown in Fig. 1(a) was prepared by photolithography on a Si chip. Both ends of the wire were connected to microstrip lines with impedance matched to 50 $\Omega$. A sinusoidal ac test signal generated by a function generator (SDG2082X, Siglent) was fed to one end and the other end was terminated by a 51 $\Omega$ chip resistor. The baseband I/Q signals were generated by an arbitrary wave generator (33622A, Keysight) with a bandwidth of 120 MHz and a sampling frequency of 1 GS/s. The baseband I and Q pulses were mixed with a microwave from a local oscillator (SMC100A, Rhodes-Schwarz) with a double-balanced mixer (IQ-1545, Marki). The amplified microwave pulses were fed to a microwave planar ring antenna~\cite{Sasaki16} with a center hole of 1 mm diameter placed above the diamond chip. NV centers were excited by a pulse laser diode at the wavelength of 520 nm. 
The photoluminescence from NV centers was imaged by a home-made wide-field microscope equipped with a scientific CMOS camera (Zyla5.5, Andor) and an objective lens $100\times$ with an NA 0.73. A pulse sequencer (Pulse Blaster ESR Pro, Spincore) controls the timings of the laser diode, the scientific  CMOS camera, the arbitrary wave generator, and the RF function generator. A static magnetic field was applied by a pair of permanent magnets parallel to the [111] direction. More details of the experiment can be found elsewhere~\cite{Nomura21,SpringerHybrid}.

\begin{figure}
\includegraphics[width=80mm]{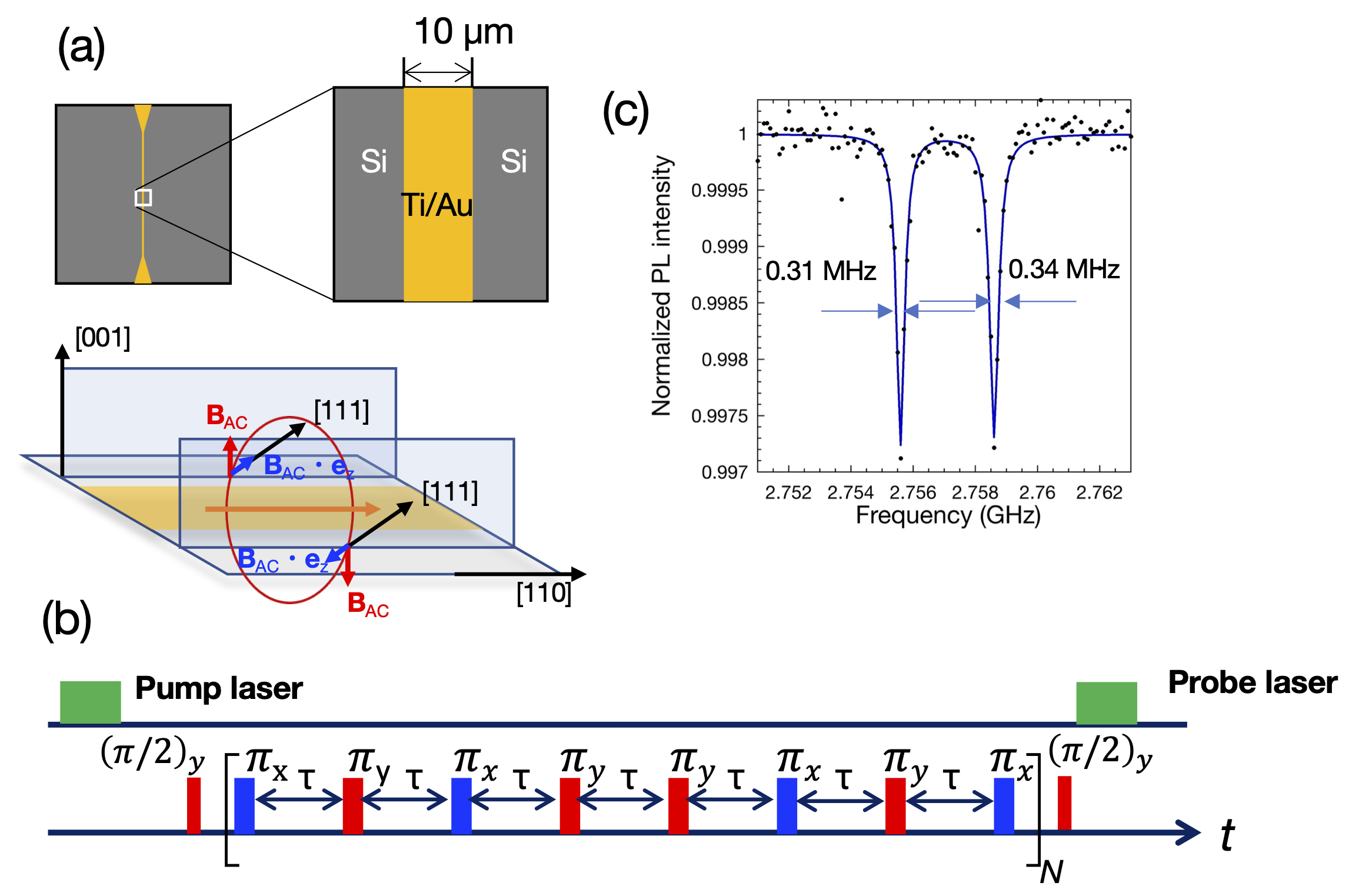}
\caption{(a) Schematics of a Ti/Au wire on a Si substrate with a width of 10 $\mu$m, and direction of the ac magnetic field ($\bf{B}_{\rm{ac}}$) produced by the current flowing in the Ti/Au wire. The ac magnetic field, $\bf{B}_{\rm{ac}}\cdot \bf{e}_{z}$, is imaged by an XY8 method, where $\bf{e}_{z}$ is a unit vector pointing in the [111] direction. (b) Pulse sequence for XY8-$N$.  $\tau$ is the period of the  $\pi$ pulses.
(c) Optically detected magnetic resonance (ODMR) spectrum of a $^{15}$N-doped diamond chip.
}
%%\label{f1}
\end{figure}

Cosine-square profile-shaped microwave pulses~\cite{Zopes17} were used in the measurements for the XY8 pulse sequence (Fig. 1(b)). 
The introduction of the shaped pulse has two advantages: first, the pulse width and pulse interval can be changed in steps much smaller than the sampling rate of an arbitrary wave generator. The high vertical resolution of the arbitrary wave generator enables us to interpolate the center position of a control microwave pulse with a timing resolution much smaller than the sampling rate of the arbitrary wave generator~\cite{Zopes17}. A time increment of less than 1 ps is possible with a 16-bit-vertical resolution and the sampling rate of 1 ns at 1 GSa/s of our setup. The second advantage is that the required bandwidth of the baseband of the pulses is kept small, and the accuracy of the spin manipulation increases. In the case of pulse control using a microwave switch, the control microwave pulse is rectangular that occupies a wide bandwidth. The second advantage is particularly important in using short control microwave pulses at a limited bandwidth of a setup.

\begin{figure}
\includegraphics[width=80mm]{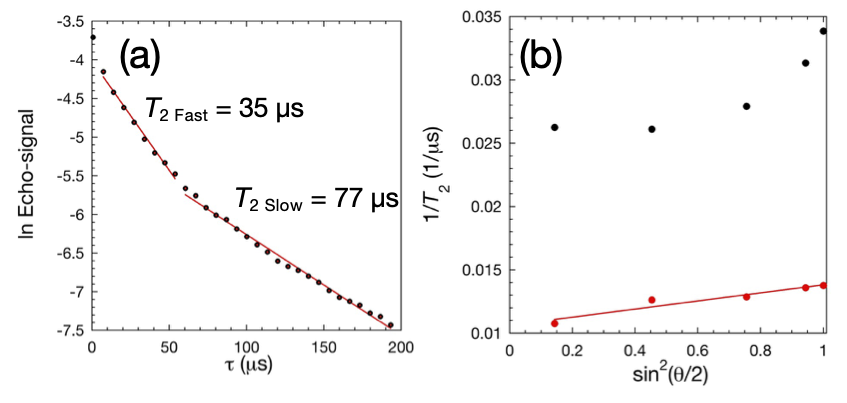}
\caption{(a) Logarithm plot of Hahn-echo coherence time $T_2$ decay profile, where $\tau$ is the time delay between the two $\pi$/2 pulses. (b) Inverse coherence time $T_{2{\rm fast}}$ and $T_{2{\rm slow}}$
as a function of ${\rm sin}^{2}\left(\frac{\theta}{2}\right)$.}
%%\label{f2}
\end{figure}

\section{Results}

Figure 1(c) shows an optically detected magnetic resonance (ODMR) spectrum of the $^{15}$N-doped diamond chip. Two structures with an energy splitting of 3.0 MHz are well-separated due to the hyperfine coupling to the $^{15}$N nuclear spin. 
The full widths at half maximum of the fitted Lorentz line profiles are 0.31 and 0.34 MHz at 2.7556 and 2.7586 GHz, respectively.
A Hahn-echo decay profile of the $^{15}$N-doped diamond chip is shown in Fig. 2(a). The obtained decay profile fits well by a double exponential function with $T_{2{\rm fast}}$ = 33 $\mu$s and $T_{2{\rm slow}}$ = 77 $\mu$s. 
Here, we note that the Hahn-echo signal decay profiles of the as-grown diamond sample and the sample after He ion injection were fitted well to a single exponential function with $T_{2}$ = 35 $\mu$s and 18 $\mu$s, respectively, which excludes the possibility of the inhomogeneous N dopant distribution as a source of the double exponential decay profile in Fig. 2(a).
An NVH$^{-}$ center is typically observed in the CVD diamond chips, but the contribution of the NVH$^{-}$ center to $T_{2}$ was found to be significantly smaller than that of N$^{0}$s, the NV$^{0}$ center, or the NV$^{-}$ center.~\cite{Chikara22}  It was also found that the concentration of the NVH$^{-}$ center before and after 1000$^{\circ}$C annealing was nearly the same.~\cite{Chikara22}

The NV density was estimated from the NV-NV spin interactions probed by instantaneous diffusion. The Hahn-echo signal is given by
\begin{equation}
S\left( n_{NV}, \theta, \tau \right)\propto
 {\rm exp }\left[-\frac{A}{4}\gamma_{NV}^{2} n_{NV}\left({\rm sin}^{2}\left(\frac{\theta}{2}\right)\right) \tau \right]
\end{equation}
where $n_{NV}$ is the density of the NV centers and and $\theta$ is the flip angle of the central pulse of the Hahn-echo sequence.~\cite{Eichhorn19}.
This model is valid in the regime where the NV-NV interaction dominates spin decoherence. The obtained Hahn-echo signals as a function of 
${\rm sin}^{2}\left(\frac{\theta}{2}\right)$
are shown in Fig. 2(b) for the fast and slow components of the decay. 
The slow component was found to be proportional to 
${\rm sin}^{2}\left(\frac{\theta}{2}\right)$
 and we obtained an NV density of 0.05 ppm.
A linear relationship to 
${\rm sin}^{2}\left(\frac{\theta}{2}\right)$
was not obtained for the fast component. 
This indicates that the initial fast decay time is not explained by the NV-NV dipolar interactions, and hence is probably due to other extrinsic factors.
In single NVs, it is generally known that the Hahn-echo signal fits well with a single exponential function. In ensemble NVs, variations in $T_2$ were observed due to variations in the couplings to the spin bath~\cite{Bauch20}. The initial fast decay is probably because the average behavior of many NVs is detected.
The long $T_{2}$ after annealing at 1000 $^{\circ}$C was previously attributed essentially due to a reduction of paramagnetic vacancy clusters near NV$^{-}$.~\cite{Yamamoto13} 
We suspect that inhomogeneity in the distribution of the paramagnetic vacancy cluster is a possible source of the double exponential decay profile.  
Fast decay depending on extrinsic effects was observed in an NMR measurement~\cite{Sasaki20}

Figure 3(a) shows a spectrum of a sinusoidal ac test signal at $f_s$ = 19.23 MHz fed to the Ti/Au metal wire structure obtained in the area of 4 $\mu$m$^{2}$ as schematically shown in Fig. 1(a) to sample the local maxima of the XY8 signals produced by the current flowing in the Ti/Au wire. The difference between the signals with and without the ac test signal is plotted. The XY8-16 sequence was used with the $\pi$ pulse length of 12.5 ns. 
A peak due to the ac test signal was observed at $\tau$ = 1/($2f_s$) = 26 ns.
The time resolution was set to be 100 ps, giving 74 kHz frequency resolution. 
This may be compared with the frequency resolution of 2.5 MHz in the case where rectangular pulses with a time resolution of 2 ns directly from the pulse sequencer were used instead of the shaped pulses. 

The detected RF frequency was higher than the previous result of 9.746 969 MHz~\cite{Zopes17} due to the shorter $\pi$ pulse length.
We repeated the measurements 10 times to obtain the XY8-16 signals similar to Fig. 3. The standard deviation ($\sigma$) at the peak position $\tau$ = 26.0 ns calculated from ten traces was 50 nT. 

An image map of the RF amplitude distribution produced by the ac current in the  Ti/Au wire structure is shown in Fig. 3(b) at $\tau$ = 26.0 ns with a pixel size of about 1 $\mu$m $\times$ 1 $\mu$m. The RF magnetic field projected to the [111] direction is detected by the XY8 protocol we used.~\cite{Childress2010,Mizuno20} The configuration of the Ti/Au wire and the crystallographic axis of the diamond chip is schematically shown in Fig. 1(a). The obtained image shows minima and maxima in the vicinity of the edges of the Ti/Au wire structure, where the ac magnetic field projected to the [111] direction is large, while the detected RF amplitude is close to zero at the center of the wire structure, where the ac magnetic field is perpendicular to the [111] direction.

\begin{figure}
\includegraphics[width=80mm]{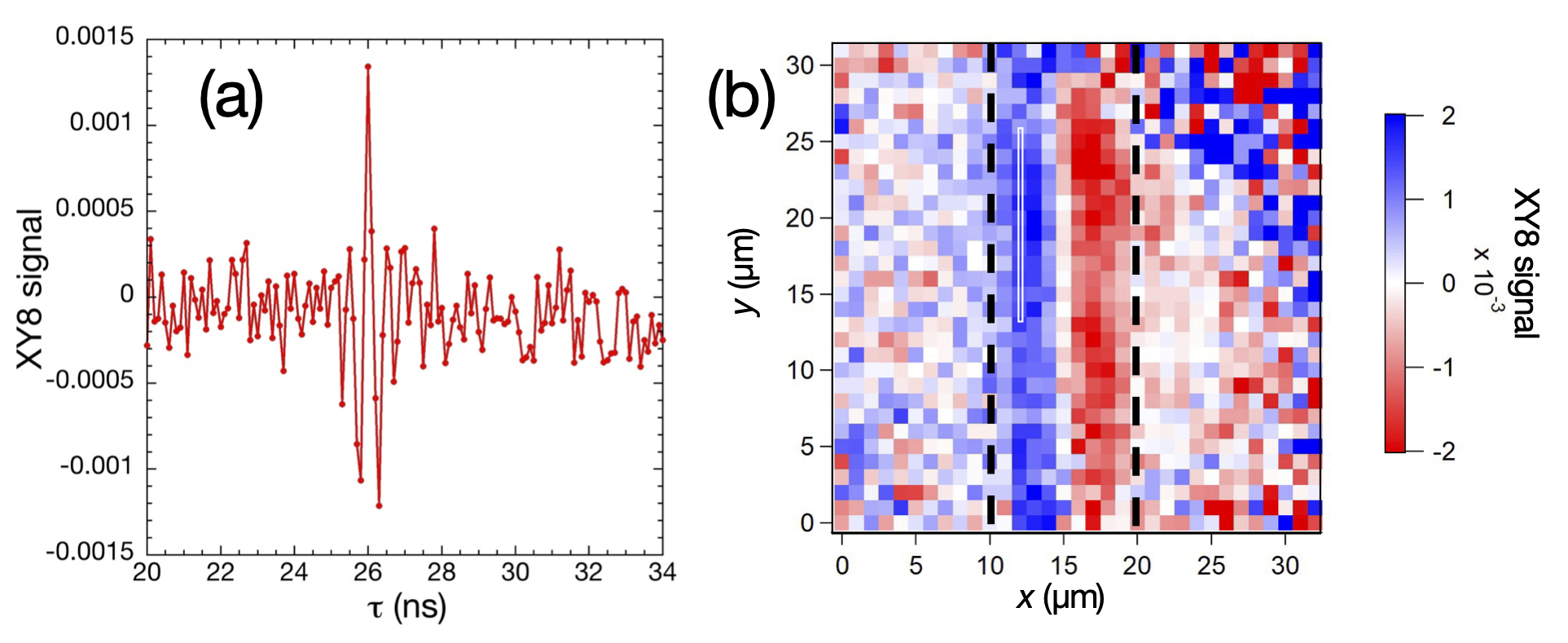}
\caption{(a) XY8-16 signal obtained by applying RF fields at 19.23 MHz and 0.44 $\mu$T. The XY8 signal integrated in the square with corners at (12.5 $\mu$m, 12.8 $\mu$m) - (12.8 $\mu$m, 25.6 $\mu$m) with the area of 4 $\mu $m$^{2}$ as indicated by the white square in (b). The exposure time was 54 ms/frame. 100 frames were averaged per data point. (b) Mapping of the RF field amplitude component projected to the [111] direction. Pixels were binned by $16\times 16$. The black dashed lines indicate the edges of the Ti/Au wire structure. }
%%\label{f3}
\end{figure}

\section{Discussion}

In this paper, we have demonstrated that RF imaging at 19.23 MHz by a multipulse dynamic decoupling method with high spectral resolution using the microwave pulses at the Rabi frequency of 40 MHz. The detected RF frequency corresponds to the $^1$H NMR frequency at 0.45 T. 
The microwave pulses at the Rabi frequency of 95.3 MHz were previously achieved in our set-up.~\cite{Giacomo20} 
The maximum frequency of the RF electromagnetic field to be detected is currently limited by the sampling frequency at 1 GS/s and the bandwidth of our arbitrary wave generator of 120 MHz. The range of the RF is expected to be extended to about a half of the available microwave Rabi frequency by using an arbitrary wave generator with higher sampling rate and wider bandwidth. 
The sensitivity may be further enhanced by increasing the number of $\pi$ pulses and optimizing the diamond chip design. 
Recently, a sensing protocol has been developed where the frequency resolution of a dynamical decoupling method is only limited by the stability of the synchronization clock.~\cite{Schmitt17, Boss17, Glenn2018} This method is suitable for applications requiring particularly high-frequency resolution and has also been applied to imaging.~\cite{Mizuno20}
This technique may allow us to detect frequency shift of $\sim$ppm range at several 10s of MHz region for NMR. Pulse control errors due to amplitude or frequency variations of the microwave pulses may be reduced by introducing the pulse shape optimization method~\cite{Rembold20}. We consider that practical high-spatial-resolution high-resolution NMR is possible by using a diamond NV center sensor.

\begin{acknowledgments}
This work was partly supported by a Grant-in-Aid for Scientific Research (Nos. 21H01009 and 22K18710) from Japan Society for the Promotion of Science. 
\end{acknowledgments}

%%\bibliography{diaxy8.bib}

\begin{thebibliography}{36}%
\makeatletter
\providecommand \@ifxundefined [1]{%
 \@ifx{#1\undefined}
}%
\providecommand \@ifnum [1]{%
 \ifnum #1\expandafter \@firstoftwo
 \else \expandafter \@secondoftwo
 \fi
}%
\providecommand \@ifx [1]{%
 \ifx #1\expandafter \@firstoftwo
 \else \expandafter \@secondoftwo
 \fi
}%
\providecommand \natexlab [1]{#1}%
\providecommand \enquote  [1]{``#1''}%
\providecommand \bibnamefont  [1]{#1}%
\providecommand \bibfnamefont [1]{#1}%
\providecommand \citenamefont [1]{#1}%
\providecommand \href@noop [0]{\@secondoftwo}%
\providecommand \href [0]{\begingroup \@sanitize@url \@href}%
\providecommand \@href[1]{\@@startlink{#1}\@@href}%
\providecommand \@@href[1]{\endgroup#1\@@endlink}%
\providecommand \@sanitize@url [0]{\catcode `\\12\catcode `\$12\catcode
  `\&12\catcode `\#12\catcode `\^12\catcode `\_12\catcode `\%12\relax}%
\providecommand \@@startlink[1]{}%
\providecommand \@@endlink[0]{}%
\providecommand \url  [0]{\begingroup\@sanitize@url \@url }%
\providecommand \@url [1]{\endgroup\@href {#1}{\urlprefix }}%
\providecommand \urlprefix  [0]{URL }%
\providecommand \Eprint [0]{\href }%
\providecommand \doibase [0]{https://doi.org/}%
\providecommand \selectlanguage [0]{\@gobble}%
\providecommand \bibinfo  [0]{\@secondoftwo}%
\providecommand \bibfield  [0]{\@secondoftwo}%
\providecommand \translation [1]{[#1]}%
\providecommand \BibitemOpen [0]{}%
\providecommand \bibitemStop [0]{}%
\providecommand \bibitemNoStop [0]{.\EOS\space}%
\providecommand \EOS [0]{\spacefactor3000\relax}%
\providecommand \BibitemShut  [1]{\csname bibitem#1\endcsname}%
\let\auto@bib@innerbib\@empty
%</preamble>
\bibitem [{\citenamefont {Zoughi}(2000)}]{Zoughi00}%
  \BibitemOpen
  \bibfield  {author} {\bibinfo {author} {\bibfnamefont {R.}~\bibnamefont
  {Zoughi}},\ }\href@noop {} {\emph {\bibinfo {title} {Microwave
  non-destructive testing and evaluation}}}\ (\bibinfo  {publisher} {Kluwer
  Academic Publishers},\ \bibinfo {address} {Dordrecht},\ \bibinfo {year}
  {2000})\BibitemShut {NoStop}%
\bibitem [{\citenamefont {Chipaux}\ \emph {et~al.}(2015)\citenamefont
  {Chipaux}, \citenamefont {Toraille}, \citenamefont {Larat}, \citenamefont
  {Morvan}, \citenamefont {Pezzagna}, \citenamefont {Meijer},\ and\
  \citenamefont {Debuisschert}}]{Chipaux15}%
  \BibitemOpen
  \bibfield  {author} {\bibinfo {author} {\bibfnamefont {M.}~\bibnamefont
  {Chipaux}}, \bibinfo {author} {\bibfnamefont {L.}~\bibnamefont {Toraille}},
  \bibinfo {author} {\bibfnamefont {C.}~\bibnamefont {Larat}}, \bibinfo
  {author} {\bibfnamefont {L.}~\bibnamefont {Morvan}}, \bibinfo {author}
  {\bibfnamefont {S.}~\bibnamefont {Pezzagna}}, \bibinfo {author}
  {\bibfnamefont {J.}~\bibnamefont {Meijer}},\ and\ \bibinfo {author}
  {\bibfnamefont {T.}~\bibnamefont {Debuisschert}},\ }\bibfield  {title}
  {\bibinfo {title} {{Wide bandwidth instantaneous radio frequency spectrum
  analyzer based on nitrogen vacancy centers in diamond}},\ }\bibfield
  {journal} {\bibinfo  {journal} {Appl. Physs Lett.}\ }\textbf {\bibinfo
  {volume} {107}},\ \href {https://doi.org/10.1063/1.4936758}
  {10.1063/1.4936758} (\bibinfo {year} {2015}),\ \bibinfo {note}
  {233502}\BibitemShut {NoStop}%
\bibitem [{\citenamefont {Horsley}\ \emph {et~al.}(2018)\citenamefont
  {Horsley}, \citenamefont {Appel}, \citenamefont {Wolters}, \citenamefont
  {Achard}, \citenamefont {Tallaire}, \citenamefont {Maletinsky},\ and\
  \citenamefont {Treutlein}}]{Horsley18}%
  \BibitemOpen
  \bibfield  {author} {\bibinfo {author} {\bibfnamefont {A.}~\bibnamefont
  {Horsley}}, \bibinfo {author} {\bibfnamefont {P.}~\bibnamefont {Appel}},
  \bibinfo {author} {\bibfnamefont {J.}~\bibnamefont {Wolters}}, \bibinfo
  {author} {\bibfnamefont {J.}~\bibnamefont {Achard}}, \bibinfo {author}
  {\bibfnamefont {A.}~\bibnamefont {Tallaire}}, \bibinfo {author}
  {\bibfnamefont {P.}~\bibnamefont {Maletinsky}},\ and\ \bibinfo {author}
  {\bibfnamefont {P.}~\bibnamefont {Treutlein}},\ }\bibfield  {title} {\bibinfo
  {title} {Microwave device characterization using a widefield diamond
  microscope},\ }\href@noop {} {\bibfield  {journal} {\bibinfo  {journal}
  {Phys. Rev. Appl.}\ }\textbf {\bibinfo {volume} {10}},\ \bibinfo {pages}
  {044039} (\bibinfo {year} {2018})}\BibitemShut {NoStop}%
\bibitem [{\citenamefont {Yang}\ \emph {et~al.}(2018)\citenamefont {Yang},
  \citenamefont {Du}, \citenamefont {Dong}, \citenamefont {Liu}, \citenamefont
  {Hu},\ and\ \citenamefont {Wang}}]{Yang18}%
  \BibitemOpen
  \bibfield  {author} {\bibinfo {author} {\bibfnamefont {B.}~\bibnamefont
  {Yang}}, \bibinfo {author} {\bibfnamefont {G.}~\bibnamefont {Du}}, \bibinfo
  {author} {\bibfnamefont {Y.}~\bibnamefont {Dong}}, \bibinfo {author}
  {\bibfnamefont {G.}~\bibnamefont {Liu}}, \bibinfo {author} {\bibfnamefont
  {Z.}~\bibnamefont {Hu}},\ and\ \bibinfo {author} {\bibfnamefont
  {Y.}~\bibnamefont {Wang}},\ }\bibfield  {title} {\bibinfo {title}
  {Non-invasive imaging method of microwave near field based on solid state
  quantum sensing},\ }\href@noop {} {\bibfield  {journal} {\bibinfo  {journal}
  {IEEE Trans. Microwave Theory and Techniques}\ }\textbf {\bibinfo {volume}
  {88}},\ \bibinfo {pages} {2276 } (\bibinfo {year} {2018})}\BibitemShut
  {NoStop}%
\bibitem [{\citenamefont {Mariani}\ \emph {et~al.}(2020)\citenamefont
  {Mariani}, \citenamefont {Nomoto}, \citenamefont {Kashiwaya},\ and\
  \citenamefont {Nomura}}]{Giacomo20}%
  \BibitemOpen
  \bibfield  {author} {\bibinfo {author} {\bibfnamefont {G.}~\bibnamefont
  {Mariani}}, \bibinfo {author} {\bibfnamefont {S.}~\bibnamefont {Nomoto}},
  \bibinfo {author} {\bibfnamefont {S.}~\bibnamefont {Kashiwaya}},\ and\
  \bibinfo {author} {\bibfnamefont {S.}~\bibnamefont {Nomura}},\ }\bibfield
  {title} {\bibinfo {title} {System for the remote control and imaging of
  $\mathrm{MW}$ fields for spin manipulation in $\mathrm{NV}$ centers in
  diamond},\ }\href@noop {} {\bibfield  {journal} {\bibinfo  {journal} {Sci.
  Rep.}\ }\textbf {\bibinfo {volume} {10}},\ \bibinfo {pages} {4813} (\bibinfo
  {year} {2020})}\BibitemShut {NoStop}%
\bibitem [{\citenamefont {Hu}\ \emph {et~al.}(2019)\citenamefont {Hu},
  \citenamefont {Yang}, \citenamefont {Dong}, \citenamefont {Liu},
  \citenamefont {Wang},\ and\ \citenamefont {Du}}]{Hu19}%
  \BibitemOpen
  \bibfield  {author} {\bibinfo {author} {\bibfnamefont {Z.}~\bibnamefont
  {Hu}}, \bibinfo {author} {\bibfnamefont {B.}~\bibnamefont {Yang}}, \bibinfo
  {author} {\bibfnamefont {M.}~\bibnamefont {Dong}}, \bibinfo {author}
  {\bibfnamefont {Y.}~\bibnamefont {Liu}}, \bibinfo {author} {\bibfnamefont
  {Y.}~\bibnamefont {Wang}},\ and\ \bibinfo {author} {\bibfnamefont
  {G.}~\bibnamefont {Du}},\ }\bibfield  {title} {\bibinfo {title} {Optical
  sensing of broadband rf magnetic field using a micrometer-sized diamond},\
  }\href {https://doi.org/10.1109/TMAG.2018.2886162} {\bibfield  {journal}
  {\bibinfo  {journal} {IEEE Trans. Magnetics}\ }\textbf {\bibinfo {volume}
  {55}},\ \bibinfo {pages} {1} (\bibinfo {year} {2019})}\BibitemShut {NoStop}%
\bibitem [{\citenamefont {Mizuno}\ \emph {et~al.}(2020)\citenamefont {Mizuno},
  \citenamefont {Ishiwata}, \citenamefont {Masuyama}, \citenamefont {Iwasaki},\
  and\ \citenamefont {Hatano}}]{Mizuno20}%
  \BibitemOpen
  \bibfield  {author} {\bibinfo {author} {\bibfnamefont {K.}~\bibnamefont
  {Mizuno}}, \bibinfo {author} {\bibfnamefont {H.}~\bibnamefont {Ishiwata}},
  \bibinfo {author} {\bibfnamefont {Y.}~\bibnamefont {Masuyama}}, \bibinfo
  {author} {\bibfnamefont {T.}~\bibnamefont {Iwasaki}},\ and\ \bibinfo {author}
  {\bibfnamefont {M.}~\bibnamefont {Hatano}},\ }\bibfield  {title} {\bibinfo
  {title} {Simultaneous wide-field imaging of phase and magnitude of ac
  magnetic signal using diamond quantum magnetometry},\ }\href
  {https://doi.org/10.1038/s41598-020-68404-5} {\bibfield  {journal} {\bibinfo
  {journal} {Sci. Rep.}\ }\textbf {\bibinfo {volume} {10}},\ \bibinfo {pages}
  {11611} (\bibinfo {year} {2020})}\BibitemShut {NoStop}%
\bibitem [{\citenamefont {Nomura}\ \emph {et~al.}(2021)\citenamefont {Nomura},
  \citenamefont {Kaida}, \citenamefont {Watanabe},\ and\ \citenamefont
  {Kashiwaya}}]{Nomura21}%
  \BibitemOpen
  \bibfield  {author} {\bibinfo {author} {\bibfnamefont {S.}~\bibnamefont
  {Nomura}}, \bibinfo {author} {\bibfnamefont {K.}~\bibnamefont {Kaida}},
  \bibinfo {author} {\bibfnamefont {H.}~\bibnamefont {Watanabe}},\ and\
  \bibinfo {author} {\bibfnamefont {S.}~\bibnamefont {Kashiwaya}},\ }\bibfield
  {title} {\bibinfo {title} {{Near-field radio-frequency imaging by
  spin-locking with a nitrogen-vacancy spin sensor}},\ }\bibfield  {journal}
  {\bibinfo  {journal} {J. Appl. Phys.}\ }\textbf {\bibinfo {volume} {130}},\
  \href {https://doi.org/10.1063/5.0052161} {10.1063/5.0052161} (\bibinfo
  {year} {2021}),\ \bibinfo {note} {024503}\BibitemShut {NoStop}%
\bibitem [{\citenamefont {Loretz}\ \emph {et~al.}(2013)\citenamefont {Loretz},
  \citenamefont {Rosskopf},\ and\ \citenamefont {Degen}}]{Loretz13}%
  \BibitemOpen
  \bibfield  {author} {\bibinfo {author} {\bibfnamefont {M.}~\bibnamefont
  {Loretz}}, \bibinfo {author} {\bibfnamefont {T.}~\bibnamefont {Rosskopf}},\
  and\ \bibinfo {author} {\bibfnamefont {C.~L.}\ \bibnamefont {Degen}},\
  }\bibfield  {title} {\bibinfo {title} {Radio-frequency magnetometry using a
  single electron spin},\ }\href@noop {} {\bibfield  {journal} {\bibinfo
  {journal} {Phys. Rev. Lett.}\ }\textbf {\bibinfo {volume} {110}},\ \bibinfo
  {pages} {017602} (\bibinfo {year} {2013})}\BibitemShut {NoStop}%
\bibitem [{\citenamefont {Ohashi}\ \emph {et~al.}(2017)\citenamefont {Ohashi},
  \citenamefont {Rosskopf}, \citenamefont {Watanabe}, \citenamefont {Loretz},
  \citenamefont {Tao}, \citenamefont {Hauert}, \citenamefont {Tomizawa},
  \citenamefont {Ishikawa}, \citenamefont {Ishi-Hayase}, \citenamefont
  {Shikata}, \citenamefont {Degen},\ and\ \citenamefont {Itoh}}]{Ohashi13}%
  \BibitemOpen
  \bibfield  {author} {\bibinfo {author} {\bibfnamefont {K.}~\bibnamefont
  {Ohashi}}, \bibinfo {author} {\bibfnamefont {T.}~\bibnamefont {Rosskopf}},
  \bibinfo {author} {\bibfnamefont {H.}~\bibnamefont {Watanabe}}, \bibinfo
  {author} {\bibfnamefont {M.}~\bibnamefont {Loretz}}, \bibinfo {author}
  {\bibfnamefont {Y.}~\bibnamefont {Tao}}, \bibinfo {author} {\bibfnamefont
  {R.}~\bibnamefont {Hauert}}, \bibinfo {author} {\bibfnamefont
  {S.}~\bibnamefont {Tomizawa}}, \bibinfo {author} {\bibfnamefont
  {T.}~\bibnamefont {Ishikawa}}, \bibinfo {author} {\bibfnamefont
  {J.}~\bibnamefont {Ishi-Hayase}}, \bibinfo {author} {\bibfnamefont
  {S.}~\bibnamefont {Shikata}}, \bibinfo {author} {\bibfnamefont {C.~L.}\
  \bibnamefont {Degen}},\ and\ \bibinfo {author} {\bibfnamefont {K.~M.}\
  \bibnamefont {Itoh}},\ }\bibfield  {title} {\bibinfo {title} {Negatively
  charged nitrogen-vacancy centers in a 5 nm thin 12c diamond film},\ }\href
  {https://doi.org/doi: 10.1021/nl402286v} {\bibfield  {journal} {\bibinfo
  {journal} {Nano Lett.}\ }\textbf {\bibinfo {volume} {13}},\ \bibinfo {pages}
  {1530} (\bibinfo {year} {2017})}\BibitemShut {NoStop}%
\bibitem [{\citenamefont {Rosskopf}\ \emph {et~al.}(2014)\citenamefont
  {Rosskopf}, \citenamefont {Dussaux}, \citenamefont {Ohashi}, \citenamefont
  {Loretz}, \citenamefont {Schirhagl}, \citenamefont {Watanabe}, \citenamefont
  {Shikata}, \citenamefont {Itoh},\ and\ \citenamefont {Degen}}]{Rosskopf14}%
  \BibitemOpen
  \bibfield  {author} {\bibinfo {author} {\bibfnamefont {T.}~\bibnamefont
  {Rosskopf}}, \bibinfo {author} {\bibfnamefont {A.}~\bibnamefont {Dussaux}},
  \bibinfo {author} {\bibfnamefont {K.}~\bibnamefont {Ohashi}}, \bibinfo
  {author} {\bibfnamefont {M.}~\bibnamefont {Loretz}}, \bibinfo {author}
  {\bibfnamefont {R.}~\bibnamefont {Schirhagl}}, \bibinfo {author}
  {\bibfnamefont {H.}~\bibnamefont {Watanabe}}, \bibinfo {author}
  {\bibfnamefont {S.}~\bibnamefont {Shikata}}, \bibinfo {author} {\bibfnamefont
  {K.~M.}\ \bibnamefont {Itoh}},\ and\ \bibinfo {author} {\bibfnamefont
  {C.~L.}\ \bibnamefont {Degen}},\ }\bibfield  {title} {\bibinfo {title}
  {Investigation of surface magnetic noise by shallow spins in diamond},\
  }\href@noop {} {\bibfield  {journal} {\bibinfo  {journal} {Phys. Rev. Lett.}\
  }\textbf {\bibinfo {volume} {112}},\ \bibinfo {pages} {147602} (\bibinfo
  {year} {2014})}\BibitemShut {NoStop}%
\bibitem [{\citenamefont {Fuchs}\ \emph {et~al.}(2009)\citenamefont {Fuchs},
  \citenamefont {Dobrovitski}, \citenamefont {Toyli}, \citenamefont
  {Heremans},\ and\ \citenamefont {Awschalom}}]{Fuchs09}%
  \BibitemOpen
  \bibfield  {author} {\bibinfo {author} {\bibfnamefont {G.~D.}\ \bibnamefont
  {Fuchs}}, \bibinfo {author} {\bibfnamefont {V.~V.}\ \bibnamefont
  {Dobrovitski}}, \bibinfo {author} {\bibfnamefont {D.~M.}\ \bibnamefont
  {Toyli}}, \bibinfo {author} {\bibfnamefont {F.~J.}\ \bibnamefont
  {Heremans}},\ and\ \bibinfo {author} {\bibfnamefont {D.~D.}\ \bibnamefont
  {Awschalom}},\ }\bibfield  {title} {\bibinfo {title} {Gigahertz dynamics of a
  strongly driven single quantum spin},\ }\href@noop {} {\bibfield  {journal}
  {\bibinfo  {journal} {Science}\ }\textbf {\bibinfo {volume} {326}},\ \bibinfo
  {pages} {1520} (\bibinfo {year} {2009})}\BibitemShut {NoStop}%
\bibitem [{\citenamefont {Cywi\ifmmode~\acute{n}\else \'{n}\fi{}ski}\ \emph
  {et~al.}(2008)\citenamefont {Cywi\ifmmode~\acute{n}\else \'{n}\fi{}ski},
  \citenamefont {Lutchyn}, \citenamefont {Nave},\ and\ \citenamefont
  {Das~Sarma}}]{Cywinski08}%
  \BibitemOpen
  \bibfield  {author} {\bibinfo {author} {\bibfnamefont {L.}~\bibnamefont
  {Cywi\ifmmode~\acute{n}\else \'{n}\fi{}ski}}, \bibinfo {author}
  {\bibfnamefont {R.~M.}\ \bibnamefont {Lutchyn}}, \bibinfo {author}
  {\bibfnamefont {C.~P.}\ \bibnamefont {Nave}},\ and\ \bibinfo {author}
  {\bibfnamefont {S.}~\bibnamefont {Das~Sarma}},\ }\bibfield  {title} {\bibinfo
  {title} {How to enhance dephasing time in superconducting qubits},\ }\href
  {https://doi.org/10.1103/PhysRevB.77.174509} {\bibfield  {journal} {\bibinfo
  {journal} {Phys. Rev. B}\ }\textbf {\bibinfo {volume} {77}},\ \bibinfo
  {pages} {174509} (\bibinfo {year} {2008})}\BibitemShut {NoStop}%
\bibitem [{\citenamefont {Ryan}\ \emph {et~al.}(2010)\citenamefont {Ryan},
  \citenamefont {Hodges},\ and\ \citenamefont {Cory}}]{Ryan10}%
  \BibitemOpen
  \bibfield  {author} {\bibinfo {author} {\bibfnamefont {C.~A.}\ \bibnamefont
  {Ryan}}, \bibinfo {author} {\bibfnamefont {J.~S.}\ \bibnamefont {Hodges}},\
  and\ \bibinfo {author} {\bibfnamefont {D.~G.}\ \bibnamefont {Cory}},\
  }\bibfield  {title} {\bibinfo {title} {Robust decoupling techniques to extend
  quantum coherence in diamond},\ }\href
  {https://doi.org/10.1103/PhysRevLett.105.200402} {\bibfield  {journal}
  {\bibinfo  {journal} {Phys. Rev. Lett.}\ }\textbf {\bibinfo {volume} {105}},\
  \bibinfo {pages} {200402} (\bibinfo {year} {2010})}\BibitemShut {NoStop}%
\bibitem [{\citenamefont {de~Lange}\ \emph {et~al.}(2010)\citenamefont
  {de~Lange}, \citenamefont {Wang}, \citenamefont {Riste}, \citenamefont
  {Dobrovitski},\ and\ \citenamefont {Hanson}}]{Lange10}%
  \BibitemOpen
  \bibfield  {author} {\bibinfo {author} {\bibfnamefont {G.}~\bibnamefont
  {de~Lange}}, \bibinfo {author} {\bibfnamefont {Z.~H.}\ \bibnamefont {Wang}},
  \bibinfo {author} {\bibfnamefont {D.}~\bibnamefont {Riste}}, \bibinfo
  {author} {\bibfnamefont {V.~V.}\ \bibnamefont {Dobrovitski}},\ and\ \bibinfo
  {author} {\bibfnamefont {R.}~\bibnamefont {Hanson}},\ }\bibfield  {title}
  {\bibinfo {title} {Universal dynamical decoupling of a single solid-state
  spin from a spin bath},\ }\href {https://doi.org/10.1126/science.1192739}
  {\bibfield  {journal} {\bibinfo  {journal} {Science}\ }\textbf {\bibinfo
  {volume} {330}},\ \bibinfo {pages} {60} (\bibinfo {year} {2010})}\BibitemShut
  {NoStop}%
\bibitem [{\citenamefont {Yuge}\ \emph {et~al.}(2011)\citenamefont {Yuge},
  \citenamefont {Sasaki},\ and\ \citenamefont {Hirayama}}]{Yuge11}%
  \BibitemOpen
  \bibfield  {author} {\bibinfo {author} {\bibfnamefont {T.}~\bibnamefont
  {Yuge}}, \bibinfo {author} {\bibfnamefont {S.}~\bibnamefont {Sasaki}},\ and\
  \bibinfo {author} {\bibfnamefont {Y.}~\bibnamefont {Hirayama}},\ }\bibfield
  {title} {\bibinfo {title} {Measurement of the noise spectrum using a
  multiple-pulse sequence},\ }\href
  {https://doi.org/10.1103/PhysRevLett.107.170504} {\bibfield  {journal}
  {\bibinfo  {journal} {Phys. Rev. Lett.}\ }\textbf {\bibinfo {volume} {107}},\
  \bibinfo {pages} {170504} (\bibinfo {year} {2011})}\BibitemShut {NoStop}%
\bibitem [{\citenamefont {Wang}\ \emph {et~al.}(2012)\citenamefont {Wang},
  \citenamefont {de~Lange}, \citenamefont {Rist\`e}, \citenamefont {Hanson},\
  and\ \citenamefont {Dobrovitski}}]{Wang12}%
  \BibitemOpen
  \bibfield  {author} {\bibinfo {author} {\bibfnamefont {Z.-H.}\ \bibnamefont
  {Wang}}, \bibinfo {author} {\bibfnamefont {G.}~\bibnamefont {de~Lange}},
  \bibinfo {author} {\bibfnamefont {D.}~\bibnamefont {Rist\`e}}, \bibinfo
  {author} {\bibfnamefont {R.}~\bibnamefont {Hanson}},\ and\ \bibinfo {author}
  {\bibfnamefont {V.~V.}\ \bibnamefont {Dobrovitski}},\ }\bibfield  {title}
  {\bibinfo {title} {Comparison of dynamical decoupling protocols for a
  nitrogen-vacancy center in diamond},\ }\href
  {https://doi.org/10.1103/PhysRevB.85.155204} {\bibfield  {journal} {\bibinfo
  {journal} {Phys. Rev. B}\ }\textbf {\bibinfo {volume} {85}},\ \bibinfo
  {pages} {155204} (\bibinfo {year} {2012})}\BibitemShut {NoStop}%
\bibitem [{\citenamefont {Zopes}\ \emph {et~al.}(2017)\citenamefont {Zopes},
  \citenamefont {Sasaki}, \citenamefont {Cujia}, \citenamefont {Boss},
  \citenamefont {Chang}, \citenamefont {Segawa}, \citenamefont {Itoh},\ and\
  \citenamefont {Degen}}]{Zopes17}%
  \BibitemOpen
  \bibfield  {author} {\bibinfo {author} {\bibfnamefont {J.}~\bibnamefont
  {Zopes}}, \bibinfo {author} {\bibfnamefont {K.}~\bibnamefont {Sasaki}},
  \bibinfo {author} {\bibfnamefont {K.~S.}\ \bibnamefont {Cujia}}, \bibinfo
  {author} {\bibfnamefont {J.~M.}\ \bibnamefont {Boss}}, \bibinfo {author}
  {\bibfnamefont {K.}~\bibnamefont {Chang}}, \bibinfo {author} {\bibfnamefont
  {T.~F.}\ \bibnamefont {Segawa}}, \bibinfo {author} {\bibfnamefont {K.~M.}\
  \bibnamefont {Itoh}},\ and\ \bibinfo {author} {\bibfnamefont {C.~L.}\
  \bibnamefont {Degen}},\ }\bibfield  {title} {\bibinfo {title}
  {High-resolution quantum sensing with shaped control pulses},\ }\href@noop {}
  {\bibfield  {journal} {\bibinfo  {journal} {Phys. Rev. Lett.}\ }\textbf
  {\bibinfo {volume} {119}},\ \bibinfo {pages} {260501} (\bibinfo {year}
  {2017})}\BibitemShut {NoStop}%
\bibitem [{\citenamefont {Joas}\ \emph {et~al.}(2017)\citenamefont {Joas},
  \citenamefont {Waeber}, \citenamefont {Braunbeck},\ and\ \citenamefont
  {Reinhard}}]{Joas2017}%
  \BibitemOpen
  \bibfield  {author} {\bibinfo {author} {\bibfnamefont {T.}~\bibnamefont
  {Joas}}, \bibinfo {author} {\bibfnamefont {A.~M.}\ \bibnamefont {Waeber}},
  \bibinfo {author} {\bibfnamefont {G.}~\bibnamefont {Braunbeck}},\ and\
  \bibinfo {author} {\bibfnamefont {F.}~\bibnamefont {Reinhard}},\ }\bibfield
  {title} {\bibinfo {title} {Quantum sensing of weak radio-frequency signals by
  pulsed mollow absorption spectroscopy},\ }\href@noop {} {\bibfield  {journal}
  {\bibinfo  {journal} {Nat. Commun.}\ }\textbf {\bibinfo {volume} {8}},\
  \bibinfo {pages} {964} (\bibinfo {year} {2017})}\BibitemShut {NoStop}%
\bibitem [{\citenamefont {Stark}\ \emph {et~al.}(2017)\citenamefont {Stark},
  \citenamefont {Aharon}, \citenamefont {Unden}, \citenamefont {Louzon},
  \citenamefont {Huck}, \citenamefont {Andersen},\ and\ \citenamefont
  {Jelezko}}]{Stark2017}%
  \BibitemOpen
  \bibfield  {author} {\bibinfo {author} {\bibfnamefont {A.}~\bibnamefont
  {Stark}}, \bibinfo {author} {\bibfnamefont {N.}~\bibnamefont {Aharon}},
  \bibinfo {author} {\bibfnamefont {T.}~\bibnamefont {Unden}}, \bibinfo
  {author} {\bibfnamefont {D.}~\bibnamefont {Louzon}}, \bibinfo {author}
  {\bibfnamefont {A.}~\bibnamefont {Huck}, \bibfnamefont {Alexander~Retzker}},
  \bibinfo {author} {\bibfnamefont {U.~L.}\ \bibnamefont {Andersen}},\ and\
  \bibinfo {author} {\bibfnamefont {F.}~\bibnamefont {Jelezko}},\ }\bibfield
  {title} {\bibinfo {title} {Narrow-bandwidth sensing of high-frequency fields
  with continuous dynamical decoupling},\ }\href@noop {} {\bibfield  {journal}
  {\bibinfo  {journal} {Nat. Commun.}\ }\textbf {\bibinfo {volume} {8}},\
  \bibinfo {pages} {1105} (\bibinfo {year} {2017})}\BibitemShut {NoStop}%
\bibitem [{\citenamefont {Saijo}\ \emph {et~al.}(2018)\citenamefont {Saijo},
  \citenamefont {Matsuzaki}, \citenamefont {Saito}, \citenamefont {Yamaguchi},
  \citenamefont {Hanano}, \citenamefont {Watanabe}, \citenamefont {Mizuochi},\
  and\ \citenamefont {Ishi-Hayase}}]{Saijo2018}%
  \BibitemOpen
  \bibfield  {author} {\bibinfo {author} {\bibfnamefont {S.}~\bibnamefont
  {Saijo}}, \bibinfo {author} {\bibfnamefont {Y.}~\bibnamefont {Matsuzaki}},
  \bibinfo {author} {\bibfnamefont {S.}~\bibnamefont {Saito}}, \bibinfo
  {author} {\bibfnamefont {T.}~\bibnamefont {Yamaguchi}}, \bibinfo {author}
  {\bibfnamefont {I.}~\bibnamefont {Hanano}}, \bibinfo {author} {\bibfnamefont
  {H.}~\bibnamefont {Watanabe}}, \bibinfo {author} {\bibfnamefont
  {N.}~\bibnamefont {Mizuochi}},\ and\ \bibinfo {author} {\bibfnamefont
  {J.}~\bibnamefont {Ishi-Hayase}},\ }\bibfield  {title} {\bibinfo {title} {{AC
  magnetic field sensing using continuous-wave optically detected magnetic
  resonance of nitrogen-vacancy centers in diamond}},\ }\href
  {https://doi.org/10.1063/1.5024401} {\bibfield  {journal} {\bibinfo
  {journal} {Appl. Phys. Lett.}\ }\textbf {\bibinfo {volume} {113}},\ \bibinfo
  {pages} {082405} (\bibinfo {year} {2018})}\BibitemShut {NoStop}%
\bibitem [{\citenamefont {Tashima}\ \emph {et~al.}(2019)\citenamefont
  {Tashima}, \citenamefont {Morishita},\ and\ \citenamefont
  {Mizuochi}}]{Tashima19}%
  \BibitemOpen
  \bibfield  {author} {\bibinfo {author} {\bibfnamefont {T.}~\bibnamefont
  {Tashima}}, \bibinfo {author} {\bibfnamefont {H.}~\bibnamefont {Morishita}},\
  and\ \bibinfo {author} {\bibfnamefont {N.}~\bibnamefont {Mizuochi}},\
  }\bibfield  {title} {\bibinfo {title} {Experimental demonstration of
  two-photon magnetic resonances in a single-spin system of a solid},\ }\href
  {https://doi.org/10.1103/PhysRevA.100.023801} {\bibfield  {journal} {\bibinfo
   {journal} {Phys. Rev. A}\ }\textbf {\bibinfo {volume} {100}},\ \bibinfo
  {pages} {023801} (\bibinfo {year} {2019})}\BibitemShut {NoStop}%
\bibitem [{\citenamefont {Ishikawa}\ \emph {et~al.}(2012)\citenamefont
  {Ishikawa}, \citenamefont {Fu}, \citenamefont {Santori}, \citenamefont
  {Acosta}, \citenamefont {Beausoleil}, \citenamefont {Watanabe}, \citenamefont
  {Shikata},\ and\ \citenamefont {Itoh}}]{Ishikawa12}%
  \BibitemOpen
  \bibfield  {author} {\bibinfo {author} {\bibfnamefont {T.}~\bibnamefont
  {Ishikawa}}, \bibinfo {author} {\bibfnamefont {K.-M.~C.}\ \bibnamefont {Fu}},
  \bibinfo {author} {\bibfnamefont {C.}~\bibnamefont {Santori}}, \bibinfo
  {author} {\bibfnamefont {V.~M.}\ \bibnamefont {Acosta}}, \bibinfo {author}
  {\bibfnamefont {R.~G.}\ \bibnamefont {Beausoleil}}, \bibinfo {author}
  {\bibfnamefont {H.}~\bibnamefont {Watanabe}}, \bibinfo {author}
  {\bibfnamefont {S.}~\bibnamefont {Shikata}},\ and\ \bibinfo {author}
  {\bibfnamefont {K.~M.}\ \bibnamefont {Itoh}},\ }\bibfield  {title} {\bibinfo
  {title} {Optical and spin coherence properties of nitrogen-vacancy centers
  placed in a 100 nm thick isotopically purified diamond layer},\ }\href
  {https://doi.org/10.1021/nl300350r} {\bibfield  {journal} {\bibinfo
  {journal} {Nano Letters}\ }\textbf {\bibinfo {volume} {12}},\ \bibinfo
  {pages} {2083} (\bibinfo {year} {2012})}\BibitemShut {NoStop}%
\bibitem [{\citenamefont {Kazi}\ \emph {et~al.}(2021)\citenamefont {Kazi},
  \citenamefont {Shelby}, \citenamefont {Watanabe}, \citenamefont {Itoh},
  \citenamefont {Shutthanandan}, \citenamefont {Wiggins},\ and\ \citenamefont
  {Fu}}]{Zeeshawn21}%
  \BibitemOpen
  \bibfield  {author} {\bibinfo {author} {\bibfnamefont {Z.}~\bibnamefont
  {Kazi}}, \bibinfo {author} {\bibfnamefont {I.~M.}\ \bibnamefont {Shelby}},
  \bibinfo {author} {\bibfnamefont {H.}~\bibnamefont {Watanabe}}, \bibinfo
  {author} {\bibfnamefont {K.~M.}\ \bibnamefont {Itoh}}, \bibinfo {author}
  {\bibfnamefont {V.}~\bibnamefont {Shutthanandan}}, \bibinfo {author}
  {\bibfnamefont {P.~A.}\ \bibnamefont {Wiggins}},\ and\ \bibinfo {author}
  {\bibfnamefont {K.-M.~C.}\ \bibnamefont {Fu}},\ }\bibfield  {title} {\bibinfo
  {title} {Wide-field dynamic magnetic microscopy using double-double quantum
  driving of a diamond defect ensemble},\ }\href
  {https://doi.org/10.1103/PhysRevApplied.15.054032} {\bibfield  {journal}
  {\bibinfo  {journal} {Phys. Rev. Appl.}\ }\textbf {\bibinfo {volume} {15}},\
  \bibinfo {pages} {054032} (\bibinfo {year} {2021})}\BibitemShut {NoStop}%
\bibitem [{\citenamefont {Yamamoto}\ \emph {et~al.}(2013)\citenamefont
  {Yamamoto}, \citenamefont {Umeda}, \citenamefont {Watanabe}, \citenamefont
  {Onoda}, \citenamefont {Markham}, \citenamefont {Twitchen}, \citenamefont
  {Naydenov}, \citenamefont {McGuinness}, \citenamefont {Teraji}, \citenamefont
  {Koizumi}, \citenamefont {Dolde}, \citenamefont {Fedder}, \citenamefont
  {Honert}, \citenamefont {Wrachtrup}, \citenamefont {Ohshima}, \citenamefont
  {Jelezko},\ and\ \citenamefont {Isoya}}]{Yamamoto13}%
  \BibitemOpen
  \bibfield  {author} {\bibinfo {author} {\bibfnamefont {T.}~\bibnamefont
  {Yamamoto}}, \bibinfo {author} {\bibfnamefont {T.}~\bibnamefont {Umeda}},
  \bibinfo {author} {\bibfnamefont {K.}~\bibnamefont {Watanabe}}, \bibinfo
  {author} {\bibfnamefont {S.}~\bibnamefont {Onoda}}, \bibinfo {author}
  {\bibfnamefont {M.~L.}\ \bibnamefont {Markham}}, \bibinfo {author}
  {\bibfnamefont {D.~J.}\ \bibnamefont {Twitchen}}, \bibinfo {author}
  {\bibfnamefont {B.}~\bibnamefont {Naydenov}}, \bibinfo {author}
  {\bibfnamefont {L.~P.}\ \bibnamefont {McGuinness}}, \bibinfo {author}
  {\bibfnamefont {T.}~\bibnamefont {Teraji}}, \bibinfo {author} {\bibfnamefont
  {S.}~\bibnamefont {Koizumi}}, \bibinfo {author} {\bibfnamefont
  {F.}~\bibnamefont {Dolde}}, \bibinfo {author} {\bibfnamefont
  {H.}~\bibnamefont {Fedder}}, \bibinfo {author} {\bibfnamefont
  {J.}~\bibnamefont {Honert}}, \bibinfo {author} {\bibfnamefont
  {J.}~\bibnamefont {Wrachtrup}}, \bibinfo {author} {\bibfnamefont
  {T.}~\bibnamefont {Ohshima}}, \bibinfo {author} {\bibfnamefont
  {F.}~\bibnamefont {Jelezko}},\ and\ \bibinfo {author} {\bibfnamefont
  {J.}~\bibnamefont {Isoya}},\ }\bibfield  {title} {\bibinfo {title} {Extending
  spin coherence times of diamond qubits by high-temperature annealing},\
  }\href {https://doi.org/10.1103/PhysRevB.88.075206} {\bibfield  {journal}
  {\bibinfo  {journal} {Phys. Rev. B}\ }\textbf {\bibinfo {volume} {88}},\
  \bibinfo {pages} {075206} (\bibinfo {year} {2013})}\BibitemShut {NoStop}%
\bibitem [{\citenamefont {Sasaki}\ \emph {et~al.}(2016)\citenamefont {Sasaki},
  \citenamefont {Monnai}, \citenamefont {Saijo}, \citenamefont {Fujita},
  \citenamefont {Watanabe}, \citenamefont {Ishi-Hayase}, \citenamefont {Itoh},\
  and\ \citenamefont {Abe}}]{Sasaki16}%
  \BibitemOpen
  \bibfield  {author} {\bibinfo {author} {\bibfnamefont {K.}~\bibnamefont
  {Sasaki}}, \bibinfo {author} {\bibfnamefont {Y.}~\bibnamefont {Monnai}},
  \bibinfo {author} {\bibfnamefont {S.}~\bibnamefont {Saijo}}, \bibinfo
  {author} {\bibfnamefont {R.}~\bibnamefont {Fujita}}, \bibinfo {author}
  {\bibfnamefont {H.}~\bibnamefont {Watanabe}}, \bibinfo {author}
  {\bibfnamefont {J.}~\bibnamefont {Ishi-Hayase}}, \bibinfo {author}
  {\bibfnamefont {K.~M.}\ \bibnamefont {Itoh}},\ and\ \bibinfo {author}
  {\bibfnamefont {E.}~\bibnamefont {Abe}},\ }\bibfield  {title} {\bibinfo
  {title} {Broadband, large-area microwave antenna for optically detected
  magnetic resonance of nitrogen-vacancy centers in diamond},\ }\href@noop {}
  {\bibfield  {journal} {\bibinfo  {journal} {Rev. Sci. Instrum.}\ }\textbf
  {\bibinfo {volume} {87}},\ \bibinfo {pages} {053904} (\bibinfo {year}
  {2016})}\BibitemShut {NoStop}%
\bibitem [{\citenamefont {Nomura}(2021)}]{SpringerHybrid}%
  \BibitemOpen
  \bibfield  {author} {\bibinfo {author} {\bibfnamefont {S.}~\bibnamefont
  {Nomura}},\ }\bibinfo {title} {Hybrid quantum systems}\ (\bibinfo
  {publisher} {Springer Nature},\ \bibinfo {year} {2021})\ Chap.~\bibinfo
  {chapter} {2}, p.~\bibinfo {pages} {27}\BibitemShut {NoStop}%
\bibitem [{\citenamefont {Shinei}\ \emph {et~al.}(2022)\citenamefont {Shinei},
  \citenamefont {Masuyama}, \citenamefont {Miyakawa}, \citenamefont {Abe},
  \citenamefont {Ishii}, \citenamefont {Saiki}, \citenamefont {Onoda},
  \citenamefont {Taniguchi}, \citenamefont {Ohshima},\ and\ \citenamefont
  {Teraji}}]{Chikara22}%
  \BibitemOpen
  \bibfield  {author} {\bibinfo {author} {\bibfnamefont {C.}~\bibnamefont
  {Shinei}}, \bibinfo {author} {\bibfnamefont {Y.}~\bibnamefont {Masuyama}},
  \bibinfo {author} {\bibfnamefont {M.}~\bibnamefont {Miyakawa}}, \bibinfo
  {author} {\bibfnamefont {H.}~\bibnamefont {Abe}}, \bibinfo {author}
  {\bibfnamefont {S.}~\bibnamefont {Ishii}}, \bibinfo {author} {\bibfnamefont
  {S.}~\bibnamefont {Saiki}}, \bibinfo {author} {\bibfnamefont
  {S.}~\bibnamefont {Onoda}}, \bibinfo {author} {\bibfnamefont
  {T.}~\bibnamefont {Taniguchi}}, \bibinfo {author} {\bibfnamefont
  {T.}~\bibnamefont {Ohshima}},\ and\ \bibinfo {author} {\bibfnamefont
  {T.}~\bibnamefont {Teraji}},\ }\bibfield  {title} {\bibinfo {title}
  {{Nitrogen related paramagnetic defects: Decoherence source of ensemble of
  NV- center}},\ }\href@noop {} {\bibfield  {journal} {\bibinfo  {journal} {J.
  of Appl. Phys.}\ }\textbf {\bibinfo {volume} {132}},\ \bibinfo {pages}
  {214402} (\bibinfo {year} {2022})}\BibitemShut {NoStop}%
\bibitem [{\citenamefont {Eichhorn}\ \emph {et~al.}(2019)\citenamefont
  {Eichhorn}, \citenamefont {McLellan},\ and\ \citenamefont
  {Bleszynski~Jayich}}]{Eichhorn19}%
  \BibitemOpen
  \bibfield  {author} {\bibinfo {author} {\bibfnamefont {T.~R.}\ \bibnamefont
  {Eichhorn}}, \bibinfo {author} {\bibfnamefont {C.~A.}\ \bibnamefont
  {McLellan}},\ and\ \bibinfo {author} {\bibfnamefont {A.~C.}\ \bibnamefont
  {Bleszynski~Jayich}},\ }\bibfield  {title} {\bibinfo {title} {Optimizing the
  formation of depth-confined nitrogen vacancy center spin ensembles in diamond
  for quantum sensing},\ }\href
  {https://doi.org/10.1103/PhysRevMaterials.3.113802} {\bibfield  {journal}
  {\bibinfo  {journal} {Phys. Rev. Mater.}\ }\textbf {\bibinfo {volume} {3}},\
  \bibinfo {pages} {113802} (\bibinfo {year} {2019})}\BibitemShut {NoStop}%
\bibitem [{\citenamefont {Bauch}\ \emph {et~al.}(2020)\citenamefont {Bauch},
  \citenamefont {Singh}, \citenamefont {Lee}, \citenamefont {Hart},
  \citenamefont {Schloss}, \citenamefont {Turner}, \citenamefont {Barry},
  \citenamefont {Pham}, \citenamefont {Bar-Gill}, \citenamefont {Yelin},\ and\
  \citenamefont {Walsworth}}]{Bauch20}%
  \BibitemOpen
  \bibfield  {author} {\bibinfo {author} {\bibfnamefont {E.}~\bibnamefont
  {Bauch}}, \bibinfo {author} {\bibfnamefont {S.}~\bibnamefont {Singh}},
  \bibinfo {author} {\bibfnamefont {J.}~\bibnamefont {Lee}}, \bibinfo {author}
  {\bibfnamefont {C.~A.}\ \bibnamefont {Hart}}, \bibinfo {author}
  {\bibfnamefont {J.~M.}\ \bibnamefont {Schloss}}, \bibinfo {author}
  {\bibfnamefont {M.~J.}\ \bibnamefont {Turner}}, \bibinfo {author}
  {\bibfnamefont {J.~F.}\ \bibnamefont {Barry}}, \bibinfo {author}
  {\bibfnamefont {L.~M.}\ \bibnamefont {Pham}}, \bibinfo {author}
  {\bibfnamefont {N.}~\bibnamefont {Bar-Gill}}, \bibinfo {author}
  {\bibfnamefont {S.~F.}\ \bibnamefont {Yelin}},\ and\ \bibinfo {author}
  {\bibfnamefont {R.~L.}\ \bibnamefont {Walsworth}},\ }\bibfield  {title}
  {\bibinfo {title} {Decoherence of ensembles of nitrogen-vacancy centers in
  diamond},\ }\href {https://doi.org/10.1103/PhysRevB.102.134210} {\bibfield
  {journal} {\bibinfo  {journal} {Phys. Rev. B}\ }\textbf {\bibinfo {volume}
  {102}},\ \bibinfo {pages} {134210} (\bibinfo {year} {2020})}\BibitemShut
  {NoStop}%
\bibitem [{\citenamefont {Sasaki}\ \emph {et~al.}(2020)\citenamefont {Sasaki},
  \citenamefont {Miura}, \citenamefont {Ikeda}, \citenamefont {Sakai},
  \citenamefont {Sekikawa}, \citenamefont {Saito}, \citenamefont {Yuge},\ and\
  \citenamefont {Hirayama}}]{Sasaki20}%
  \BibitemOpen
  \bibfield  {author} {\bibinfo {author} {\bibfnamefont {S.}~\bibnamefont
  {Sasaki}}, \bibinfo {author} {\bibfnamefont {T.}~\bibnamefont {Miura}},
  \bibinfo {author} {\bibfnamefont {K.}~\bibnamefont {Ikeda}}, \bibinfo
  {author} {\bibfnamefont {M.}~\bibnamefont {Sakai}}, \bibinfo {author}
  {\bibfnamefont {T.}~\bibnamefont {Sekikawa}}, \bibinfo {author}
  {\bibfnamefont {M.}~\bibnamefont {Saito}}, \bibinfo {author} {\bibfnamefont
  {T.}~\bibnamefont {Yuge}},\ and\ \bibinfo {author} {\bibfnamefont
  {Y.}~\bibnamefont {Hirayama}},\ }\bibfield  {title} {\bibinfo {title}
  {$1/f^{2}$ spectra of decoherence noise on $^{75} \mathrm{As}$ nuclear spins
  in bulk $\mathrm{GaAs}$},\ }\href@noop {} {\bibfield  {journal} {\bibinfo
  {journal} {Sci. Rep.}\ }\textbf {\bibinfo {volume} {10}},\ \bibinfo {pages}
  {10674} (\bibinfo {year} {2020})}\BibitemShut {NoStop}%
\bibitem [{\citenamefont {Childress}\ and\ \citenamefont
  {McIntyre}(2010)}]{Childress2010}%
  \BibitemOpen
  \bibfield  {author} {\bibinfo {author} {\bibfnamefont {L.}~\bibnamefont
  {Childress}}\ and\ \bibinfo {author} {\bibfnamefont {J.}~\bibnamefont
  {McIntyre}},\ }\bibfield  {title} {\bibinfo {title} {Multifrequency spin
  resonance in diamond},\ }\href {https://doi.org/10.1103/PhysRevA.82.033839}
  {\bibfield  {journal} {\bibinfo  {journal} {Phys. Rev. A}\ }\textbf {\bibinfo
  {volume} {82}},\ \bibinfo {pages} {033839} (\bibinfo {year}
  {2010})}\BibitemShut {NoStop}%
\bibitem [{\citenamefont {Schmitt}\ \emph {et~al.}(2017)\citenamefont
  {Schmitt}, \citenamefont {Gefen}, \citenamefont {Stürner}, \citenamefont
  {Unden}, \citenamefont {Wolff}, \citenamefont {Müller}, \citenamefont
  {Scheuer}, \citenamefont {Naydenov}, \citenamefont {Markham}, \citenamefont
  {Pezzagna}, \citenamefont {Meijer}, \citenamefont {Schwarz}, \citenamefont
  {Plenio}, \citenamefont {Retzker}, \citenamefont {McGuinness},\ and\
  \citenamefont {Jelezko}}]{Schmitt17}%
  \BibitemOpen
  \bibfield  {author} {\bibinfo {author} {\bibfnamefont {S.}~\bibnamefont
  {Schmitt}}, \bibinfo {author} {\bibfnamefont {T.}~\bibnamefont {Gefen}},
  \bibinfo {author} {\bibfnamefont {F.~M.}\ \bibnamefont {Stürner}}, \bibinfo
  {author} {\bibfnamefont {T.}~\bibnamefont {Unden}}, \bibinfo {author}
  {\bibfnamefont {G.}~\bibnamefont {Wolff}}, \bibinfo {author} {\bibfnamefont
  {C.}~\bibnamefont {Müller}}, \bibinfo {author} {\bibfnamefont
  {J.}~\bibnamefont {Scheuer}}, \bibinfo {author} {\bibfnamefont
  {B.}~\bibnamefont {Naydenov}}, \bibinfo {author} {\bibfnamefont
  {M.}~\bibnamefont {Markham}}, \bibinfo {author} {\bibfnamefont
  {S.}~\bibnamefont {Pezzagna}}, \bibinfo {author} {\bibfnamefont
  {J.}~\bibnamefont {Meijer}}, \bibinfo {author} {\bibfnamefont
  {I.}~\bibnamefont {Schwarz}}, \bibinfo {author} {\bibfnamefont
  {M.}~\bibnamefont {Plenio}}, \bibinfo {author} {\bibfnamefont
  {A.}~\bibnamefont {Retzker}}, \bibinfo {author} {\bibfnamefont {L.~P.}\
  \bibnamefont {McGuinness}},\ and\ \bibinfo {author} {\bibfnamefont
  {F.}~\bibnamefont {Jelezko}},\ }\bibfield  {title} {\bibinfo {title}
  {Submillihertz magnetic spectroscopy performed with a nanoscale quantum
  sensor},\ }\href {https://doi.org/10.1126/science.aam5532} {\bibfield
  {journal} {\bibinfo  {journal} {Science}\ }\textbf {\bibinfo {volume}
  {356}},\ \bibinfo {pages} {832} (\bibinfo {year} {2017})}\BibitemShut
  {NoStop}%
\bibitem [{\citenamefont {Boss}\ \emph {et~al.}(2017)\citenamefont {Boss},
  \citenamefont {Cujia}, \citenamefont {Zopes},\ and\ \citenamefont
  {Degen}}]{Boss17}%
  \BibitemOpen
  \bibfield  {author} {\bibinfo {author} {\bibfnamefont {J.~M.}\ \bibnamefont
  {Boss}}, \bibinfo {author} {\bibfnamefont {K.~S.}\ \bibnamefont {Cujia}},
  \bibinfo {author} {\bibfnamefont {J.}~\bibnamefont {Zopes}},\ and\ \bibinfo
  {author} {\bibfnamefont {C.~L.}\ \bibnamefont {Degen}},\ }\bibfield  {title}
  {\bibinfo {title} {Quantum sensing with arbitrary frequency resolution},\
  }\href {https://doi.org/10.1126/science.aam7009} {\bibfield  {journal}
  {\bibinfo  {journal} {Science}\ }\textbf {\bibinfo {volume} {356}},\ \bibinfo
  {pages} {837} (\bibinfo {year} {2017})}\BibitemShut {NoStop}%
\bibitem [{\citenamefont {Glenn}\ \emph {et~al.}(2018)\citenamefont {Glenn},
  \citenamefont {Bucher}, \citenamefont {Lee}, \citenamefont {Lukin},
  \citenamefont {Park},\ and\ \citenamefont {Walsworth}}]{Glenn2018}%
  \BibitemOpen
  \bibfield  {author} {\bibinfo {author} {\bibfnamefont {D.~R.}\ \bibnamefont
  {Glenn}}, \bibinfo {author} {\bibfnamefont {D.~B.}\ \bibnamefont {Bucher}},
  \bibinfo {author} {\bibfnamefont {J.}~\bibnamefont {Lee}}, \bibinfo {author}
  {\bibfnamefont {M.~D.}\ \bibnamefont {Lukin}}, \bibinfo {author}
  {\bibfnamefont {H.}~\bibnamefont {Park}},\ and\ \bibinfo {author}
  {\bibfnamefont {R.~L.}\ \bibnamefont {Walsworth}},\ }\bibfield  {title}
  {\bibinfo {title} {High-resolution nuclear magnetic resonance spectroscopy at
  the scale of single cells is achieved by combining a magnetometer consisting
  of an ensemble of nitrogen-vacancy centres with a narrowband synchronized
  readout protocol},\ }\href {https://doi.org/10.1038/nature25781} {\bibfield
  {journal} {\bibinfo  {journal} {Nature}\ }\textbf {\bibinfo {volume} {555}},\
  \bibinfo {pages} {351} (\bibinfo {year} {2018})}\BibitemShut {NoStop}%
\bibitem [{\citenamefont {Rembold}\ \emph {et~al.}(2020)\citenamefont
  {Rembold}, \citenamefont {Oshnik}, \citenamefont {Muller}, \citenamefont
  {Montangero}, \citenamefont {Calarco},\ and\ \citenamefont
  {Neu}}]{Rembold20}%
  \BibitemOpen
  \bibfield  {author} {\bibinfo {author} {\bibfnamefont {P.}~\bibnamefont
  {Rembold}}, \bibinfo {author} {\bibfnamefont {N.}~\bibnamefont {Oshnik}},
  \bibinfo {author} {\bibfnamefont {M.~M.}\ \bibnamefont {Muller}}, \bibinfo
  {author} {\bibfnamefont {S.}~\bibnamefont {Montangero}}, \bibinfo {author}
  {\bibfnamefont {T.}~\bibnamefont {Calarco}},\ and\ \bibinfo {author}
  {\bibfnamefont {E.}~\bibnamefont {Neu}},\ }\bibfield  {title} {\bibinfo
  {title} {{Introduction to quantum optimal control for quantum sensing with
  nitrogen-vacancy centers in diamond}},\ }\bibfield  {journal} {\bibinfo
  {journal} {AVS Quantum Science}\ }\textbf {\bibinfo {volume} {2}},\ \href
  {https://doi.org/10.1116/5.0006785} {10.1116/5.0006785} (\bibinfo {year}
  {2020}),\ \bibinfo {note} {024701}\BibitemShut {NoStop}%
\end{thebibliography}

\end{document}